\documentclass
[12pt]{article}
\usepackage{amssymb}
\usepackage{epsfig}

\catcode`\@=11
\def\numberbysection{\@addtoreset{equation}{section}
        \def\theequation{\thesection.\arabic{equation}}}

\def\beq{\begin{equation}}
\def\eeq{\end{equation}}
\numberbysection
\begin{document}
\begin{titlepage}
\begin{center}
\hfill  \\

\vskip 1.in {\Large \bf Exactly solvable $f(R)$ inflation} \vskip 0.5in P. Valtancoli
\\[.2in]
{\em Dipartimento di Fisica, Polo Scientifico Universit\'a di Firenze \\
and INFN, Sezione di Firenze (Italy)\\
Via G. Sansone 1, 50019 Sesto Fiorentino, Italy}
\end{center}
\vskip .5in
\begin{abstract}
We show that adding a cosmological constant term to the Starobinsky model, it can be solved exactly without using the slow-roll approximation.
\end{abstract}
\medskip
\end{titlepage}
\pagenumbering{arabic}
\section{Introduction}

To explain the homogeneity, flatness and horizon problems of the universe it has been hypothesized the existence of a rapid expansion phase of the primordial universe. Inflationary cosmology can also predict the anisotropies of the cosmic background temperature and the formation of large-scale structures.
The first models of inflation were based on the coupling of one or more scalar fields ( the inflaton ) to standard gravity and correspond to a modification of the energy-momentum tensor in Einstein's equations.

However, there is another approach to explain the accelerated expansion of the universe. This corresponds to modified gravity in which the gravitational theory is modified with respect to general relativity. A subclass of such models is the gravity $ f(R) $ in which the lagrangian density is an arbitrary function of $ R $ \cite{1}-\cite{2}-\cite{3}. The model $ f(R) \ = \ R \ + \ \alpha \ R^2 $ with $ \alpha > 0 $ can lead to the accelerated expansion of the universe due to the presence of the term $ \alpha \ R^2 $ , as proposed by Starobisnky in the $ 1980 $. In all cases, the slow-roll approximation is used to extract physical information from these models. Our aim is to go beyond this approximation ( see also \cite{4} ) and try to solve exactly the Starobinsky model. Our trick is to add a cosmological constant term to the model, which doesn't alter its main predictions and allows us to find an exact solution of the Friedmann equations.

Our work is organized as follows. First, the Friedmann equations of the modified Starobinsky model are resolved in the Jordan frame and the temporal dependence of all the physical quantities of the background is calculated exactly. Then we map this solution to the Einstein frame and the analysis of the physical quantities is repeated in full detail. Finally, for completeness we recall how to connect the Starobinsky model to the experimental data.

\section{$f(R)$ theory}

Consider a $f(R)$ theory in the flat space-time $FLRW$ with the metric tensor in the form $ds^2 \ = \ - dt^2 + a^2 (t) \ dx_i^2 $. We work in the notation $ k^2 \ = \ 8 \pi G \ = \ 1 $. The background metric satisfies the following Friedmann equations (in the absence of matter and radiation)

\begin{eqnarray}
3 F H^2 & = & \frac{F R - f}{2} \ - \ 3 H  \dot{F} \nonumber \\
- 2 F \dot{H} & = & \ddot{F} \ - \ H  \dot{F}
 \label{21} \end{eqnarray}

where we introduced the notation

\beq F(R) \ = \ \frac{d f(R)}{d R} \label{22}
\eeq

It is well known that the $f(R)$ gravity can be expressed as a Brans-Dicke theory in the Jordan frame. In fact the action of the model $f(R)$ can be represented as follows:

\beq  S \ = \ \int \ d^4 x \sqrt{-g} \left[ \  \frac{1}{2} \varphi R \ - \ U( \varphi ) \ \right] \label{23}
\eeq

where

\beq \varphi \ = \ F(R) \ \ \ \ \ \ \ \ U ( \varphi ) \ = \ \frac{F R - f}{2} \label{24} \eeq

The auxiliary field $\varphi$ is coupled in a non-minimal way to gravity (Jordan frame). The Friedmann equations can be rewritten as:

\begin{eqnarray}
& \ & \ddot \varphi \ + \ 3 H \dot \varphi \ + \frac{2}{3} ( \ \varphi U_\varphi \ - \ 2 U \ ) \ = \ 0 \nonumber \\
& \ & 3 \left( \ H  + \frac{\dot\varphi}{2\varphi} \ \right)^2 \ = \ \frac{3}{4} \ \frac{\dot\varphi^2}{\varphi^2} \ + \ \frac{U}{\varphi}
 \label{25}
\end{eqnarray}

where $U_\varphi = \frac{d U}{d \varphi}$. The amount $U(\varphi)$ represents an energy density, but can not be connected to an effective force. The effective potential is instead

\beq U_{eff} \ = \ \int \ \frac{2}{3} ( \ \varphi U_\varphi - 2 U \ ) d \varphi \ \ \ \ \ \ \ \ \ F_{eff} \ = \ \frac{d U_{eff} }{d \varphi}
\label{26} \eeq

In the Jordan frame we try to solve exactly the Starobinsky model:

\beq
f(R) \ = \ R \ + \ \frac{R^2}{6 M^2}
\label{27} \eeq

Our contribution is to note that by adding a cosmological constant term to (\ref{27}) we obtain a model which can be solved exactly:

\beq
f(R) \ = \ R \ + \ \frac{R^2}{6 M^2} \ + \ \frac{2 M^2}{3}
\label{28}
\eeq

The exact solution of the model (\ref{28}) that we want to analyze here is of the type:

\beq
H(t) \ = \ \frac{M}{2} \ \left( \ \sqrt{\varphi (t)} \ - \ \frac{1}{3 \ \sqrt{\varphi (t)} } \
\right)
\label{29}
\eeq

where the time dependency of $\varphi(t)$ is set to be:

\beq
\dot{\varphi} (t) \ = \ - \frac{2}{3} M \sqrt{\varphi (t)}  \ \ \ \ \ \ \ \ \ddot{\varphi} (t) \ = \ \frac{2}{9} \ M^2
\label{210}
\eeq

In turn, the time $t$ can be rewritten in terms of the $\varphi$ variable (in inflation the $\varphi$ field is homogeneous):

\beq
t \ = \ - \frac{3}{M} \ \sqrt{\varphi (t)}
\label{211}
\eeq

This solution allows to calculate the following quantities:

\begin{eqnarray}
U( \varphi ) & = & \frac{1}{2} [ R F(R) - f(R) ] \ = \ \frac{3}{4} \ M^2 \ {( \varphi - 1 )}^2  \ - \ \frac{M^2}{3}
\nonumber \\
R & = & 6 \ ( \dot{H} \ + \ 2 H^2 ) \ = \ 3 \ M^2 \ ( \varphi - 1 )
\label{212} \end{eqnarray}

from which we can verify that the Friedmann equations (\ref{25}) are solved exactly.

So we have an inflation model $f(R)$ in which we can go beyond the slow-roll approximation and check if this is a good approximation. The solution for $H(t)$ can be re-expressed as follows:

\beq
H(t) \ = \  \frac{d}{dt} \ \ln a(t) \ = \ - \frac{3}{4} \ \dot{\varphi} (t) \ + \ \frac{1}{4} \ ( \ \dot{\ln \varphi} \ )
\label{213}
\eeq

from which we can integrate for $a(t)$

\beq
a(t) \ = \  k \ \varphi(t)^\frac{1}{4} \ e^{-\frac{3}{4} \varphi (t) }
\label{214}
\eeq

Once we know $a(t)$ we can determine the conformal time

\beq
\eta(t) \ = \  \int \ \frac{dt}{a(t)} \ = \ \frac{1}{k} \ \int^t_{t_i} \ dt \ \varphi(t)^{-\frac{1}{4}} \ e^{\frac{3}{4} \varphi (t) }
\label{215}
\eeq

which is an incomplete gamma function.

The number of $N$ e-foldings between the time $t_i$ in which inflation is triggered and the time $t_f$ in which it ends is defined by:

\beq
N \ = \  \int_{t_i}^{t_f} \ H(t) \ dt \ = \ \frac{1}{4} \ ( \ \ln ( \varphi_f ) \ - \ \ln ( \varphi_i ) \ ) \ + \ \frac{3}{4} \ ( \ \varphi_i  -  \varphi_f \ )
\label{216}
\eeq

which typically is in the order of $60-70$. We note that the logarithmic term gives a small correction (1 \%) to the main term that is obtained from the slow-roll approximation, proving that this is a good approximation.

Let's take a closer look at the solution for $a(t)$. Since $\varphi(t) = \frac{M^2 t^2}{9}$ we obtain:

\beq
a(t) \ = \  k \ ( -t )^\frac{1}{2} \ e^{ - \frac{M^2}{12} t^2 }
\label{217}
\eeq

For $t$ negative $a(t)$ can be an increasing function of $t$. Calculating the derivative of $a(t)$ we obtain that

\beq
H(t) \ = \ \frac{1}{2t} \ - \ \frac{M^2}{6} t
\label{218}
\eeq

The condition $\dot{a}(t)=0$ is reached for $ t = - \frac{\sqrt{3}}{M} $, but this time is never reached, because inflation has validity only for $ \ddot{a}(t) \ > 0 $:

\beq
\ddot{a} (t) > 0 \ \ \ \ \ \rightarrow \ \ \ \ \ \  t < t_f \ = \ - \frac{\sqrt{6 + 3 \sqrt{5} }}{M}
\label{219}
\eeq

The time $t_i$ in which inflation is triggered is related to the number of $N$ e-foldings
\beq
N \ = \ \int^{t_f}_{t_i} \ H(t) dt \ = \ \frac{1}{2} \ln ( t_f - t_i ) \ - \ \frac{M^2}{12} ( t_f^2 - t_i^2 )
\label{220}
\eeq

The logarithmic term is negligible therefore we obtain:

\beq
t^2_i \ = \ t^2_f \ + \ \frac{12 N}{M^2} \ \ \ \ \ \ \ \rightarrow \ \ \ \ \ \  t_i \ = \ - \frac{\sqrt{ 12 N + 6 + 3 \sqrt{5} }}{M}
\label{221}
\eeq

We can also link the time $t$ to the $H$ function by reversing the relation (\ref{218})

\beq
t \ = \ - \frac{3H}{M^2} \ \left[  \ 1 + \sqrt{1 + \frac{M^2}{3 H^2}}
\right]
\label{222}
\eeq

For the initial time $t_i$ we know that the field $H^2_i \gg M^2$ for which we can approximate

\beq
t_i \ \sim \ -  \frac{6 H_i}{M^2} \ \sim - \frac{\sqrt{12 N}}{M}
 \ \ \ \ \ \ \ \rightarrow \ \ \ \ \ H_i \sim \ \sqrt{\frac{N}{3}} \ M
\label{223}
\eeq

while the time $t_f$ is tied to the $H_f$ field

\beq
H_f \sim \frac{M}{\sqrt{6}}
\label{224}
\eeq

which is the value for $H_f$ that is obtained from the slow-roll approximation.

Finally, we can link the number of $N$ e-foldings to the $H_i$ field

\beq N \sim \frac{M^2}{12} \ ( t_i^2 - t_f^2 ) \sim \frac{3 H_i}{M^2} \sim \frac{1}{2 \epsilon_1 ( t_i )}
\label{225} \eeq

where we introduced the slow-roll parameter

\beq
\epsilon_1 \ = \ - \frac{\dot{H}}{H^2} \ \ \ \ \ \ \epsilon_1 (t_i) \sim \frac{M^2}{6 H_i^2}
\label{226} \eeq

Later it will be useful to introduce other slow-roll parameters

\beq
\epsilon_1 \ = \ - \frac{\dot{H}}{H^2} \ \ \ \ \ \ \epsilon_2 \ = \ 0 \ \ \ \ \ \epsilon_3 \ = \ \frac{\dot{F}}{2 H F} \ \ \ \ \ \ \ \epsilon_4 \ = \ \frac{\ddot{F}}{ H \dot{F}}
\label{227} \eeq

The $\epsilon_2$ parameter is null for a theory of pure $f(R)$ inflation ( see \cite{1} ). The following exact identity is valid:

\beq
\epsilon_1 \ = \ - \epsilon_3 \ ( 1 - \epsilon_4 )
\label{228} \eeq

In the model (\ref{28})

\beq
\epsilon_3 \ = \ \epsilon_4 \ = \ \frac{2}{1 - \frac{M^2}{3}t^2} < 0 \ \ \ \ \ \ \epsilon_1 \ = \ \frac{2( 1 + \frac{M^2}{3}t^2 ) }{( 1 - \frac{M^2}{3}t^2)^2} > 0
\label{229}  \eeq

Note that for $ t \sim t_i $ the parameters $ \epsilon_1, \epsilon_3, \epsilon_4 $ are all much less than $1$,

\beq
\epsilon_1 (t_i) \ \sim \ - \ \epsilon_3 (t_i) \sim \frac{M^2}{6 H_i^2} \ll 1
\label{230} \eeq

while for $ t = t_f $ the same parameters are of the order of the unity.

\section{Einstein frame}

In the Jordan frame the $\varphi$ field is coupled not minimally to gravity. To pass to the Einstein frame in which the gravitational part of the action regains its canonical form, we must apply the following transformation:

\beq
\tilde{g}_{\mu\nu} \ = \ \varphi \ g_{\mu\nu} \ \ \ \ \ \ d \tau \ = \ \sqrt{\varphi (t)} \ dt \ \ \ \ \ \ \ \tilde{a} ( \tau ) \ = \ \sqrt{\varphi (t)} \ a(t)
\label{31}
\eeq

which leads to an action of the form:

\beq
S \ = \ \int \ d^4 x \ \sqrt{- \tilde{g}} \ \left[ \frac{1}{2} \widetilde{R} \ - \ \frac{3}{4} \ \left( \frac{\widetilde{\nabla} \varphi}{\varphi}
\right)^2 \ - \ \frac{U( \varphi )}{\varphi^2} \right]
\label{32}
\eeq

where $ \widetilde{\nabla} $ is the derivative with respect to $ \widetilde{x}^\mu $ ($ \tau \ = \ \tilde{t} $). To obtain the canonical kinetic term for $ \varphi $ we must introduce the scalar field $ \phi $ of the Einstein frame

\beq
\phi \ = \ \sqrt{\frac{3}{2}} \ \ln \varphi
\label{33}
\eeq

The canonical action in terms of $ \tilde{g}_{\mu\nu} $ and $ \phi $ is then

\beq
S \ = \ \int \ d^4 x \ \sqrt{- \tilde{g}} \ \left[ \frac{1}{2} \widetilde{R} \ - \ \frac{1}{2} \ ( \widetilde{\nabla} \phi )^2 \ - \ V( \phi ) \right]
\label{34}
\eeq

where the potential $ V(\phi) $ is calculable as

\beq
V( \phi ) \ = \ \frac{( F R - f )}{2 F^2} \ = \ \frac{U ( \varphi )}{ \varphi^2 }|_{\varphi = \varphi( \phi )}
\label{35}
\eeq

The equations of motion are:

\beq
\frac{d^2 \phi}{d \tau^2} \ + \ 3 \ \widetilde{H} \ \frac{d \phi}{d \tau } \ + \ \frac{\partial V( \phi)}{\partial \phi} \ = \ 0 \ \ \ \ \ \ \ \ \widetilde{H}^2 \ = \ \frac{1}{3} \ \left[ \ \frac{1}{2} \left( \frac{d \varphi}{d \tau} \right)^2 \ + \ V ( \phi ) \ \right]
\label{36}
\eeq

Based on the mapping (\ref{31}) we obtain the following identities:

\beq
\frac{d}{d \tau} \ = \ \frac{1}{\sqrt{\varphi (t)}} \frac{d}{dt} \ = \ - \frac{2 M}{3} \ \frac{d}{d \varphi} \ = \ - \sqrt{\frac{2}{3}} M \ e^{-\sqrt{\frac{2}{3}} \phi} \ \frac{d}{d \phi}
\label{37}
\eeq

from which we derive $ \widetilde{H} = \frac{1}{\tilde{a}} \frac{d \tilde{a}}{d \tau} $:

\beq
\widetilde{H} \ = \ \frac{1}{\sqrt{F}} \ \left(  H + \frac{\dot{F}}{2 F} \right) \ = \ \frac{M}{2} \left(  \ 1 - e^{-\sqrt{\frac{2}{3}} \phi} \ \right)
\label{38}
\eeq

Its derivative with respect to $ \tau $ is

\beq
\frac{d \widetilde{H}}{d\tau} \ = \ - \frac{M^2}{3} \  e^{-2\sqrt{\frac{2}{3}} \phi}
\label{39}
\eeq

since

\beq
\frac{d \phi}{d\tau} \ = \ - \ \sqrt{\frac{2}{3}}  M \ e^{-\sqrt{\frac{2}{3}} \phi}
\label{310}
\eeq

The curvature tensor $ \widetilde{R} $ turns out to be:

\beq
\widetilde{R} \ = \ 6 \left( \frac{d \widetilde{H}}{d\tau} \ + \ 2 \widetilde{H}^2 \right)\ = \ M^2 \ ( \ 3 - 6 \ e^{-\sqrt{\frac{2}{3}} \phi} \ + \ e^{-2\sqrt{\frac{2}{3}} \phi} \ )
\label{311}
\eeq

The potential of the solvable model is of the type:

\beq
V( \phi ) \ = \ M^2 \left[ \ \frac{3}{4} \ - \ \frac{3}{2} \ e^{-\sqrt{\frac{2}{3}} \phi} \ + \ \frac{5}{12} \ e^{-2\sqrt{\frac{2}{3}} \phi} \ \right]
\label{312}
\eeq

from which we can verify that the Friedmann equations are solved exactly. The time variable $ \tau $ is bound to the $ \phi $ field from the relation (valid only for $ \tau $ negative)

\beq
\tau \ = \ - \frac{3}{2M} \ e^{\sqrt{\frac{2}{3}} \phi}
\label{313}
\eeq

from which we get:

\beq
\widetilde{H} ( \tau ) \ = \ \frac{M}{2} \ + \ \frac{3}{4\tau}
\label{314}
\eeq

By integrating with respect to $ \tau $ we finally obtain the temporal evolution of $ \tilde{a} (\tau) $:
\beq
\tilde{a} ( \tau ) \ = \ k \ {( - \tau )}^{\frac{3}{4}} \ e^{\frac{M}{2} \tau}
\label{315}
\eeq

For $ \tau $ negative the function $ \tilde{a} (\tau) $ can be an increasing function of $ \tau $. For $ \tau = - \frac{3}{2M} $ we have $ \dot{\tilde{a}} (\tau) = 0 $, but this value is never reached. We must indeed impose that:

\beq
\ddot{\tilde{a}} ( \tau ) \ > \ 0 \ \ \ \ \ \ \ \ \ \tau < \tau_f \ = \ - \ \frac{3}{2M} \ - \ \frac{\sqrt{3}}{M}
\label{316}
\eeq

Let us calculate the slow-roll approximation variables:

\begin{eqnarray}
\epsilon ( \tau ) & = & - \ \frac{\dot{\widetilde{H}}( \tau ) }{\widetilde{H}^2( \tau )} \ = \ \frac{3}{{(\tau M + \frac{3}{2})}^2} \ \ \ \ \ \ \epsilon ( \tau_f ) \ = \ 1 \nonumber \\
\eta ( \tau ) & = & \epsilon ( \tau ) \ - \ \frac{\dot{\epsilon} ( \tau )}{2 \epsilon ( \tau ) \widetilde{H} ( \tau )} \ = \ \frac{2}{(\tau M + \frac{3}{2})} \ \ \ \ \ \ \ | \eta ( \tau_f ) | \ = \ \frac{2}{\sqrt{3}} \ \sim \ 1 \label{317}
\end{eqnarray}

Finally, the value of $ \phi_f $ at the end of the inflation is:

\beq
e^{\sqrt{\frac{2}{3}} \phi_f} \ = \ 1 \ + \ \frac{2}{\sqrt{3}} \ \ \ \ \ \ \ \ \ \phi_f \ \sim \ 1
\label{318}
\eeq

Let's calculate $ \tau_i $, that is the time when inflation starts. The number of e-foldings is

\begin{eqnarray}
N & = &  \int^{\tau_f}_{\tau_i} \ \widetilde{H} ( \tau ) \ d \tau \ = \ \frac{M}{2} ( \tau_f - \tau_i ) \ + \ \frac{1}{2} \sqrt{\frac{3}{2}}( \phi_f - \phi_i ) \ = \
\nonumber \\ & = & \frac{3}{4} \left( \ e^{\sqrt{\frac{2}{3}} \phi_i} \ - \ e^{\sqrt{\frac{2}{3}} \phi_f} \ \right) \ + \ \frac{1}{2} \sqrt{\frac{3}{2}}( \phi_f - \phi_i )
\label{319}
\end{eqnarray}

The linear contribution in $ (\phi_f - \phi_i) $ is negligible compared to the exponential of $ \phi_i $, which is what we get from the slow-roll approximation. In first approximation

\beq
\frac{M}{2} ( \tau_f - \tau_i ) \ = \ N \ \ \ \ \ \ \rightarrow \ \ \ \ \ \ \tau_i \ = \ \tau_f \ - \ \frac{2N}{M}
\label{320}
\eeq

We can calculate $ \epsilon (\tau_i) $ at the time of the inflation trigger:

\beq
\epsilon ( \tau_i ) \ = \ \frac{3}{{( \sqrt{3} + 2 N )}^2} \ \ll \ 1 \ \ \ \ \ \ \  N \sim 60
\label{321}
\eeq

\section{Perturbation equations}

For the sake of completeness, let us recall how the Starobinsky model is linked to the observables. To do this it is necessary to study a perturbed metric with respect to the $ FLRW $ flat background:

\begin{eqnarray}
ds^2 & = & - \ ( 1 + 2 \alpha ) dt^2 \ - \ 2 a(t) ( \ \partial_i \beta - S_i \ ) dt dx^i \ + \ \nonumber \\
& + & a^2 (t) ( \ \delta_{ij} \ + \ 2 \psi \delta_{ij} \ + \ 2 \partial_i \partial_j \gamma \ + \ 2 \partial_i F_j \ + \ h_{ij} \ ) dx^i dx^j
\label{41}
\end{eqnarray}

where $ \alpha, \beta, \psi, \gamma $ are scalar perturbations, $ S_i, F_i $ are vector perturbations and $ h_{ij} $ are tensor perturbations. In general, vector perturbations are not important in cosmology.

Varying the action of gravity $ f(R) $ with respect to $ g_{\mu\nu} $ we obtain the field equations

\beq
F \ R_{\mu\nu} \ - \ \frac{1}{2} \ f(R) \ g_{\mu\nu} \ - \ \nabla_\mu \nabla_\nu F \ + \ g_{\mu\nu} \ \Box \ F \ = \ 0
\label{42}
\eeq

The following perturbed quantities are then defined:

\beq
\chi \ = \ a(t) \ ( \ \beta \ + \ a(t) \ \dot{\gamma} \ ) \ \ \ \ \ \ \ \ A \ = \ 3 \ ( \ H \ \alpha \ - \ \dot{\psi} \ ) \ - \ \frac{\triangle}{a^2 (t)} \ \chi
\label{43}
\eeq

By analyzing the field equations in the gauge condition $ \delta F \ = \ 0 $, the scalar perturbations $ \alpha $ and $ A $ can be expressed in terms of $ \psi $ ( \ the curvature perturbation that we will later call $ R $ \ ). At the end $ R $ satisfies the following simple equation in the Fourier space:
\beq
\ddot{R} \ + \ \frac{(\dot{a^3 Q_s})}{a^3 Q_s} \ \dot{R} \ + \ \frac{k^2}{a^2} \ R \ = \ 0
\label{44}
\eeq

where $ k $ is a comoving wavenumber and

\beq
Q_s \ = \ \frac{3 \dot{F}^2}{2 F \ [H + \frac{\dot{F}}{2F}]^2}
\label{45}
\eeq

Introducing the variables $ z_s = a \sqrt{Q_s} $ e $ u = z_s R $ we obtain

\beq
u'' \ + \ \left( \ k^2 \ - \ \frac{z_s''}{z_s} \ \right) \ u \ = \ 0
\label{46}
\eeq

where a prime represents a derivative with respect to the conformal time $ \eta \ = \ \int \ a^{-1} (t) dt $. To derive the spectrum of the curvature perturbation generated during inflation, the following variables are introduced
\beq
\epsilon_1 \ = \ - \ \frac{\dot{H}}{H^2} \ \ \ \ \ \ \ \epsilon_2 \ = \ 0 \ \ \ \ \ \ \epsilon_3 \ = \ \frac{\dot{F}}{ 2 H F} \ \ \ \ \ \ \ \epsilon_4 \ = \ \frac{\ddot{F}}{ H \dot{F}}
\label{47}
\eeq

Equation (\ref{46}) is very difficult to solve analytically. In practice we can only solve in the limit $ \dot{\epsilon}_i \ = \ 0 $ ($ \epsilon_i $ slowly varying parameters ).
For $ \epsilon_1 $ constant, the conformal time can be approximated as follows:

\beq
\eta \ = \ - \ \frac{1}{( 1 - \epsilon_1) a H }
\label{48}
\eeq

and

\beq
\frac{z_s''}{z_s} \ =  \ \frac{\nu^2_R - \frac{1}{4}}{\eta^2} \ \ \ \ \ \ \ \ \nu^2_R \ = \ \frac{1}{4} \ + \ \frac{( 1 + \epsilon_1 - \epsilon_3 + \epsilon_4 )( 2 - \epsilon_3 + \epsilon_4 )}{{(1-\epsilon_1)}^2}
\label{49}
\eeq

The solution can be expressed as a linear combination of Hankel functions $ H^{(i)}_{\nu_R} ( k | \eta | ) $. The power spectrum of the curvature perturbation is then defined
as

\beq
P_R \ = \ \frac{4 \pi k^3}{ {( 2 \pi )}^3} \ {| R |}^2
\label{410}
\eeq

which can be calculated using the solution in terms of Hankel functions $ H^{(i)}_{\nu_R} ( k | \eta | ) $.

The spectrum must be evaluated when crossing the Hubble radius $ k \ = \ aH $ and it can be estimated as:

\beq P_R \sim \frac{1}{Q_s} \ \left( \frac{H}{2 \pi} \right)^2 \sim \frac{1}{3 \pi F} \ \left( \frac{H}{m_{pl}} \right)^2 \ \frac{1}{\epsilon_1^2}
\label{411}
\eeq

We can define the spectral index of $ R $

\beq n_R \ - \ 1 \ = \ \frac{d \ln P_R}{d \ln k}|_{k = a H} \ = \ 3 \ - \ 2 \ \nu_R \sim - \ 4 \ \epsilon_1 \ + \ 2 \ \epsilon_3 \ - \ 2 \ \epsilon_4
\label{412}
\eeq

For the tensor perturbations we obtain analogous equations that lead to the following amplitude:

\beq P_T \sim \frac{16}{\pi} \ \left( \frac{H}{m_{pl}} \right)^2 \ \frac{1}{F}
\label{413}
\eeq

The tensor-to-scalar ratio is then defined as

\beq r \ = \ \frac{P_T}{P_R} \ \sim \ \frac{64 \pi}{m_{pl}^2} \ \frac{Q_s}{F} \ \sim 48 \ \epsilon_1^2
\label{414}
\eeq

In the Starobinsky model we can approximate

\beq
F \sim \frac{4 H^2}{M^2}
\label{415}
\eeq

from which we obtain

\beq P_R \sim \frac{1}{12 \pi} \ \left( \frac{M}{m_{pl}} \right)^2 \ \frac{1}{\epsilon_1^2} \ \ \ \ \ \ \ \
P_T \sim \frac{4}{\pi} \ \left( \frac{M}{m_{pl}} \right)^2
\label{416}
\eeq

$ \epsilon_1^2 $ must be evaluated at the time $ t_k $ of the Hubble radius crossing ($ k \ = \ a H $) and can be linked to the number of e-foldings from $ t = t_k $ to $ t_f $ (the end of inflation)

\beq
N_k \sim \frac{1}{2 \epsilon_1 ( t_k )}
\label{417}
\eeq

Then the amplitude of the curvature perturbation is given by

\beq P_R \sim \frac{N_k^2}{3 \pi} \ \left( \frac{M}{m_{pl}} \right)^2
\label{418}
\eeq

Comparing with the experimental data ($ N_k \sim 55 $) the mass $ M $ is constrained to be

\beq
M \sim 3 \times 10^{-6} \ m_{pl}
\label{419}
\eeq

The spectral index of $ R $ is reduced to

\beq
n_R \ - \ 1 \sim - \ 4 \ \epsilon_1 \ \sim  \ - \ \frac{2}{N_k} \ \sim \ - \ 3.6 \ \times \ 10^{-2}
\label{420}
\eeq

Finally, the tensor-to-scalar ratio can be estimated as

\beq
r \ \sim \ \frac{12}{N_k^2} \ \sim \  4.0 \ \times \ 10^{-3}
\label{421}
\eeq

These are the observables of the Starobinsky model that are related to the experiment \cite{5}.

\section{Conclusion}

In this article we have studied how to solve the Starobinsky model by adding an appropriate cosmological constant to the model. Once we know the solution of the Friedmann equations, we can go back to the exact parameterization of the background metric $ a (t) $ from which all the properties of the solvable model derive. The solution has been first studied in the Jordan frame and then generalized to the Einstein frame where a solvable model is obtained for a scalar field coupled to standard gravity. We compared the exact solution with the model statements obtained by the slow-roll approximation, finding a substantial agreement. For the sake of completeness we have then recalled how the Starobinsky model is linked to the experimental observations made with the Planck satellite \cite{5}. We hope that the proposed framework will contribute to the clarification of the theory of inflation.

\end{document}